\documentclass[%
reprint,
amsmath,amssymb,
aps,
]{revtex4-1}

\usepackage{graphicx}
\usepackage{subfloat}
\usepackage{dcolumn}
\usepackage{bm}

\begin{document}

\preprint{APS/123-QED}

\title{Quaternion Space-Time and Matter}
\author{Viktor Ariel}
\thanks{Thanks to:  Prof. L. Altschul, Prof. S. Ruschin, and V. Matizen for helpful discussions}%
\affiliation{%
}%

\begin{abstract}
In this work, we use the concept of quaternion time and demonstrate that it can be applied for description of four-dimensional space-time intervals. We demonstrate that the quaternion time interval together with the finite speed of light propagation allow for a simple intuitive understanding of the time interval measurement during arbitrary relative motion between a signal source and observer. We derive a quaternion form of Lorentz time dilation and show that the norm corresponds to the traditional expression of the Lorentz transformation and represents the measured value of time intervals, making the new theory inseparable from the theory of measurement. We determine that the space-time interval in the observer reference frame is given by a conjugate quaternion expression, which is essential for proper definition of the quaternion derivatives in the observer reference frame. Then, we apply quaternion differentiation to an arbitrary potential, which leads to generalized Lorentz force. The second quaternion derivative of the potential leads to expressions similar to generalized Maxwell equations. Finally, we apply the resulting formalism to electromagnetic and gravitational interactions and show that the new expressions are similar to the traditional expressions, with the exception of additional terms, related to scalar fields, that need further study and experimental verification. Therefore, the new mathematical approach based on Hamilton's quaternions may serve as a useful foundation of the unified theory of space-time and matter. \\
\end{abstract}

\pacs{Valid PACS appear here}
\maketitle


\section{\label{sec:level1}Introduction\protect\\}  
 
We begin by proposing the algebra of real quaternions \cite{Rodrigues}, \cite{Hamilton}, \cite{Hamilton2}, \cite{Graves}, \cite{Ariel1}, \cite{Ariel2},  as an alternative to the traditional mathematical formalism of the special relativity theory  \cite{Poincare}, \cite{Minkowski1}, \cite{Minkowski2}, \cite{Einstein}, \cite{Moriconi}. The goal is not to reproduce conceptual conclusions of the theory of relativity, but to achieve an alternative interpretation of the Lorentz time dilation \cite{Lorentz1}, \cite{Lorentz2}.

Previously, bi-quaternions were applied to special relativity \cite{Silberstein}  and showed initial promise in developing a unified field theory \cite{Gsponer}. However unlike real quaternions, bi-quaternion mathematics is not a division algebra and therefore can not be relied upon in our opinion.  

In this work, we use real quaternions, which form a division algebra giving  confidence in the validity of the resulting mathematical methods. We develop a complex polar form of the quaternion time interval and demonstrate that it describes transition time from one physical state to another. On the other hand, the norm of the quaternion time interval describes the experimentally measured value of time, which corresponds to the Lorentz time dilation. Consequently in this theory, experimental measurements are inseparable  from the theoretical prediction and form a single coherent formulation of physical experience. 

We deduce that the conjugate quaternion time interval corresponds to the time interval in the observer reference frame, which is essential for the correct definition of quaternion differentiation in the observer reference frame.

Jack \cite{PM Jack1}, \cite{PM Jack2} demonstrated an approach of applying quaternion differentiation to derive Maxwell equations, then,  Dunning-Davies and Norman \cite{Dunning-Davies} suggested using a similar method for the gravitational field. Application of electromagnetic analogy to gravitational fields and forces was previously extensively studied \cite{Heaviside}, \cite{Jefimenko}, \cite{Mashhoon},  \cite{Fedosin}.

Therefore in this work, we introduce quaternion differentiation of a generalized quaternion potential in order to derive the generalized Lorentz force field expressions. Applying the second quaternion differentiation leads us to the unified expressions for the electromagnetic and gravitational matter density.

It appears that applying quaternion calculus  to electromagnetic and gravitational interactions reproduces the well-known results for the vector fields, while showing additional scalar and vector components that may lead to exciting new effects.

Thus, we try to show that the quaternion algebra sn calculus can serve as a basis for an alternative approach  to the unified theory of fields, matter, and reality \cite{Penrose}.

\section{Quaternion Space-Time} %

Historically, Rodrigues \cite{Rodrigues} introduced quaternions while searching for a method to describe rotation of three-dimensional solids. His discovery can be considered the precursor to quaternion algebra, which was formally introduced and extensively studied by Hamilton \cite{Hamilton}, \cite{Hamilton2}, who came across quaternions while searching for mathematical division in the three-dimensional space. Hamilton was quoted: "Time is said to have only one dimension, and space to have three dimensions. The mathematical quaternion partakes of both these elements" \cite{Graves}.  In Hamilton's definition of quaternions, time is a real scalar and space is a three-dimensional imaginary vector, which seems like a brilliant insight predating the discovery of the four-dimensional space-time.

The key advantages of real quaternion algebra over other mathematical methods are: a positive Euclidean norm, description of both rotation and propagation in three-dimensional space, and  well-defined division.  Consequently, quaternion algebra deserves further investigation as an alternative mathematical formalism of space-time physics.

Since the algebra of real quaternions is the only four-dimensional division algebra, we introduce the four-dimensional quaternion manifold,

\begin{equation}
\boldsymbol {\tau}^4
=\left ( \thinspace {\hat \tau_0} \thinspace,  \vec \tau_1, \vec \tau_2 , \vec \tau_3 \right ) 
=\left ( \thinspace {\hat \imath_0} {\tau}_0 \thinspace,  {{\vec  \imath_1} \tau_1}, {{\vec  \imath_2} \tau_2} ,  { {\vec  \imath_3} \tau_3} \right )   ,
\label{eq:qt_quaternion_time_definition}
\end{equation}
which we identify with time \cite{Ariel1} in order to facilitate an intuitive physical interpretation.

Here, $\hat \imath_0$, is a real unity scalar and,  $ \vec\imath_1, \vec\imath_2, \vec\imath_3$, are purely imaginary unit vectors, and  $ \tau_0, \tau_1,\tau_2, \tau_3  \in \mathbb R$, are real scalars. The relationships between the Euclidean quaternion units, $\hat \imath_0, \vec\imath_1, \vec\imath_2, \vec\imath_3$, are essential for the present theory and are defined according to Hamilton \cite{Hamilton} as,
\begin{equation}
\begin{cases}
{ \hat\imath}_0 \thinspace {\hat \imath}_0  = 
{ \hat \imath}_0 =  1\thinspace,
\\
\\
\vec { \imath}_1 \thinspace \vec  { \imath}_1
= \vec  { \imath}_2\thinspace  \vec {\imath}_2
= \vec { \imath}_3 \thinspace \vec  {\imath}_3
= \vec  { \imath}_1\thinspace  \vec {\imath}_2 \thinspace 
\vec  { \imath}_3
= -  { \imath}_0 = -1 \thinspace,
\\
\\
\vec {\imath}_1 \thinspace \vec {\imath}_2
= \vec { \imath}_3, \quad
\vec  {\imath}_2 \thinspace \vec  {\imath}_3
= \vec  { \imath}_1, \quad
\vec {\imath}_3 \thinspace \vec {\imath}_1
= \vec  {\imath}_2\thinspace,
\\
\\
\vec {\imath}_2 \thinspace \vec {\imath}_1
= -\vec { \imath}_3, \quad
\vec  {\imath}_3 \thinspace \vec  {\imath}_2
= -\vec  { \imath}_1, \quad
\vec {\imath}_1 \thinspace \vec {\imath}_3
= -\vec  {\imath}_2\thinspace.
\end{cases}		
\label{eq:qt_unit_intervals}
\end{equation}

In the current work, we develop the quaternion formalism in vacuum, therefore, we use the absolute value of the speed of light in vacuum, $c$, as a scalar coefficient of proportionality between space and time. This allows us to express four-dimensional space-time in terms of four-dimensional quaternion time,
\begin{equation}
\boldsymbol {\tau}^4 
=\left ( \thinspace {\hat \imath_0} {\tau}_0 \thinspace, {\vec  \imath_1} \dfrac { x_1} {c}, {\vec  \imath_2} \dfrac { x_2} {c}, {\vec  \imath_3}  \dfrac  {x_3} {c} \thinspace \right ) .
\label{eq:qt_quaternion_time_define}
\end{equation}
Interestingly, this may be comparable to the modern approach to space location measurement using time-of-flight with the help of at least four satellites in the Global Positioning System.

Thus, using the quaternion definitions (\ref{eq:qt_quaternion_time_definition}), (\ref{eq:qt_unit_intervals}) and the speed of light, we express the four-dimensional space-time in terms of quaternion time. 

\section{Quaternion space-time coordinates and intervals} %

\begin{figure}
	\includegraphics{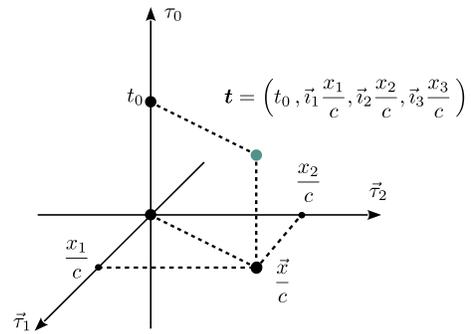}
	\caption{\label{fig:QT_Source_Coordinates} A three-dimensional representation of the quaternion time-point.}
\end{figure}

\begin{figure}
	\includegraphics{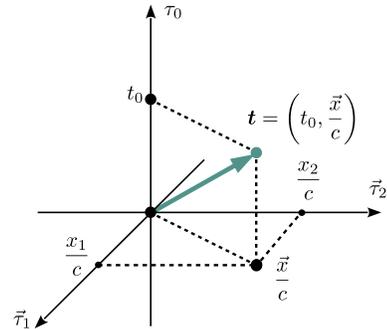}
	\caption{\label{fig:QT_Source_Interval} A three-dimensional representation of the quaternion time interval.}
\end{figure}

Next, we use quaternion space-time in order to establish coordinate locations in the four-dimensional coordinate system.

Using (\ref{eq:qt_quaternion_time_definition}) we define a point location in the quaternion space-time coordinate system as,
\begin{equation}
\boldsymbol { t}
= ( t_0, \thinspace \vec  t_v\thinspace) 
= \left ( {t_0} ,
\thinspace \dfrac {\vec  x} {c}
\thinspace \right)	
\thinspace ,
\label{eq:qt_time_coordinate_point}
\end{equation}
where $\vec x$ is a pure imaginary space vector, as original introduced by Hamilton \cite{Hamilton}
\begin{equation}
\vec x 
=\left ( \thinspace {{\vec  \imath_1} x_1} , {{\vec  \imath_2} x_2} , { {\vec  \imath_3} x_3}  \thinspace \right )
\thinspace ,
\label{eq:qt_vector_location}
\end{equation}
and, $ t_0$ is a real scalar time,
\begin{equation}
 t_0
= \hat \imath_0 t_0 = t_0
\thinspace .
\label{eq:qt_zero_point_time}
\end{equation}
Note from (\ref{eq:qt_time_coordinate_point}), that $t_0$ is the time at the zero space location $\vec x = 0$. 

The space-time coordinate point  (\ref{eq:qt_time_coordinate_point}) is defined relative to the quaternion zero-point,
\begin{equation}
\boldsymbol 0 
=\left ( \thinspace  0 \thinspace,\vec   0 \thinspace\right ) 
=\left ( \thinspace {\hat \imath_0} 0 \thinspace, {\vec  \imath_1} 0, {\vec  \imath_2} 0, {\vec  \imath_3}  0 \thinspace \right )
\thinspace .
\label{eq:qt_zero_point}
\end{equation}
Consequently, a quaternion space-time coordinate point (\ref{eq:qt_time_coordinate_point}) can be considered a four-dimensional quaternion interval starting at the zero-point and ending at the specific quaternion coordinate point defined by (\ref{eq:qt_time_coordinate_point}).

Let us consider the norm of the time interval,

\begin{equation}
{ t}
= |\boldsymbol{{  t}}|
= \sqrt{\boldsymbol{ t}\thinspace \boldsymbol{{  \bar  t}}}
=\sqrt{\boldsymbol{{\bar   t}}\thinspace  \boldsymbol{  t}}
\thinspace ,
\label{eq:qt_norm_define}
\end{equation}
where we use the conjugate quaternion time definition, 
\begin{equation}
\boldsymbol{ \bar  t}	
=\left ( \thinspace { t}_0 \thinspace, \thinspace -\vec  t_v \thinspace \thinspace\right ) 
= \left ( {t_0} ,
\thinspace- \dfrac {\vec  x} {c}
\thinspace \right)	
\thinspace .
\label{eq:qt_conj_define}
\end{equation}

In Fig.~\ref{fig:QT_Source_Coordinates} and Fig.~\ref{fig:QT_Source_Interval}, we show diagrams of a space-time point and quaternion interval using a three-dimensional representation, with a real scalar dimension, $\tau_0$, and two imaginary vector dimensions, $\vec \tau_1$ and $\vec \tau_2$. For simplicity, we neglect the fourth vector dimension, $\vec  \tau_3$.

\par
\section{Polar Representation of Quaternion Time Intervals} %
\par

Let us assume that the quaternion time interval signifies a transition in space-time from the zero-point to a space location, $\vec x$.

To describe this motion, we introduce a vector velocity,
\begin{equation}
\vec {v}
= \dfrac {{\vec x}} {  t}  
\thinspace ,		
\label{eq:qt_vector velocity}
\end{equation}
where, $\vec x$, is a space interval and, $ t =|\boldsymbol  t|$, is the absolute value of the time interval given by (\ref{eq:qt_norm_define}). Note that previously we defined an alternative quaternion velocity expression \cite{Ariel1}.

Then, we write quaternion time in terms of its norm and vector velocity, 

\begin{equation}
\begin{cases}
\boldsymbol{  t } 
= ( t_0, \thinspace \vec  t_v\thinspace) 
=    \left ({\thinspace  { t_0}}   \thinspace ,  \dfrac {\vec v}  {c} t  \thinspace \right )  
\thinspace ,
\\
\\
\boldsymbol{ \bar t } 
= ( t_0, \thinspace -\vec  t_v\thinspace) 
=    \left ({\thinspace  { t_0}}   \thinspace ,  -\dfrac {\vec v}  {c} t  \thinspace \right )  ,
\end{cases}	
\label{eq:qt_time_velocity}
\end{equation}
where we note that the quaternion time interval consists of two components: the zero-point time, $t_0$, and a vector, $({\vec v} /c) \thinspace  t$, which looks like a high-speed correction due to motion.

Next, we introduce a purely imaginary unit-vector, 
\begin{equation}
\vec {\imath}
= \dfrac {{\vec x}} { x} 
= \dfrac {\vec v} {v} 
\thinspace ,		
\label{eq:qt_unit_directinal_vector}
\end{equation}
which signifies the direction of motion.

Finally from (\ref{eq:qt_time_velocity}) and (\ref{eq:qt_unit_directinal_vector}), we express the quaternion time interval in  polar form,
\begin{equation}
\boldsymbol{   t } 
=    t \thinspace \left (\cos\theta \thinspace, \thinspace
{\vec\imath} \thinspace  \sin \theta \right )  
=   t\thinspace \exp 
\left ( {  \vec\imath \thinspace \theta}  \right ),	
\label{eq:qt_interval_polar}
\end{equation}
where the angle, $\theta$, is a function of the velocity, $\vec v$,  and is defined as, 
\begin{equation}
\begin{cases}
\cos \theta
= \dfrac { t_0}  {  t } 
=  {\sqrt {1- \dfrac {v^2} {c^2}}}
\thinspace ,
\\
\\
\sin \theta 
= \dfrac {v} {c} .
\end{cases}	
\label{eq:qt_polar_time_components}
\end{equation}

\begin{figure}
	\includegraphics{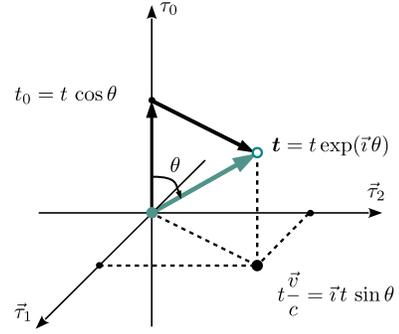}
	\caption{\label{fig:QT_Polar_Frame} Polar quaternion form of the Lorentz time interval dilation.}
\end{figure} 
\begin{figure}
	\includegraphics{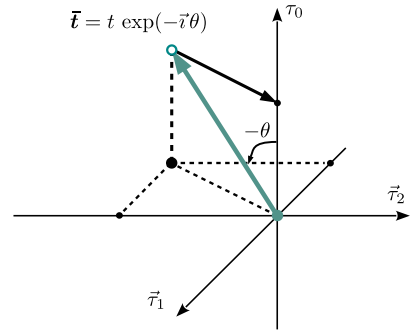}
	\caption{\label{fig:QT_Polar_Frame_Observer}  Conjugate polar form of the  Lorentz time interval dilation.}
\end{figure} 

Then from (\ref{eq:qt_interval_polar}) and (\ref{eq:qt_polar_time_components}), we obtained the full polar form  of the time interval transformation,

\begin{equation}
\begin{cases}
\boldsymbol{   t } 
=\dfrac {t_0}    {\sqrt {1- \dfrac {v^2} {c^2}}}
\thinspace \exp \left ( {  \vec\imath \thinspace \theta}  \right )  
\thinspace ,
\\
\\
\boldsymbol{\bar   t } 
=  \dfrac {   t_0 } 
{\sqrt {1- \dfrac {v^2} {c^2}}} \thinspace \exp \left ( { - \vec\imath \thinspace \theta}  \right ) \thinspace,
\end{cases}	
\label{eq:qt_quaternion_Lorentz}
\end{equation}
which we can a quaternion form of the Lorentz time dilation.

Now, we can easily determine the real scalar norm of the quaternion time interval from (\ref{eq:qt_quaternion_Lorentz}),
	
\begin{equation}
 t
= |\boldsymbol{{  t }}| = |\boldsymbol{{ \bar t }}|
=\dfrac { t_0} {\sqrt {1- \dfrac {v^2} {c^2}}} \thinspace,
\label{eq:qt_lorentz_dialation}
\end{equation}
which we immediately recognize as the traditional form of the Lorentz time dilation.

In Fig.~\ref{fig:QT_Polar_Frame} and Fig.~\ref{fig:QT_Polar_Frame_Observer}, we demonstrate diagrams of a quaternion space-time interval and its conjugate  using a three-dimensional representation.

Therefore, we realize that the time interval defined here depends on both the relative speed, $ v /c$, and direction of motion,  $\vec \imath$. On the other hand, the absolute value of the time interval is a function of the speed only.

Therefore, we were able to obtain the Lorentz time dilation by using quaternion formulation of the space-time interval and its absolute value.

\section{Physical Interpretation of Quaternion Space-Time Intervals} %

We will now elaborate on the physical meaning of the quaternion time interval defined by (\ref{eq:qt_quaternion_Lorentz}) and (\ref{eq:qt_lorentz_dialation}) .
Let us assume the existence of time sources such as clocks, and signal detectors, such as observers with recording instruments.

Assume that there is a signal source, which is a stationary clock located on a train platform. First we perform an experiment in the source reference frame of the stationary clock, where the location of the clock is the zero of space, $\vec x = 0$. Also, let us consider an observer with a video camera passing the platform on a train at midnight, when the time on the platform clock is zero. We assume that the train is moving along a straight track with a constant vector velocity, $\vec v$ . The observer synchronizes the camera timer with the platform clock at midnight and then starts filming the time on the platform clock while simultaneously recording the time-stamp of the camera. 

After synchronization, the starting time for both the platform clock and the observer camera is zero. The observer stops filming when the camera records time, $t_0$, appearing on the platform clock. What would be the time-stamp on observer's camera at the end of the recording? Due to the finite speed of light propagation, we expect that the recorded time of the platform clock will appear delayed relative to the time-stamp on the observer's camera. Also, we expect that the delay is a function of the train speed relative to the speed of light as the light signal from the clock is chasing the observer on the moving train.

Let us suggest that the time interval in the source reference frame can be defined by a quaternion, 
\begin{equation}
\boldsymbol{ t } 
=\left ( \thinspace t_0 \thinspace, \thinspace   \dfrac  {\vec v} {c} t\thinspace \right ) 
\thinspace ,
\label{eq:qt_remote_time_interval}
\end{equation}
as presented in Fig.~\ref{fig:QT_Polar_Frame}.

Also, let us suggest that the measured time interval on the camera time-stamp is a real scalar value, equal to the quaternion norm of the interval (\ref{eq:qt_lorentz_dialation}),
\begin{equation}
|\boldsymbol{{  t }}| 
= \sqrt{\boldsymbol{ t}\thinspace \boldsymbol{{  \bar  t}}}
=\sqrt{\boldsymbol{{\bar   t}}\thinspace  \boldsymbol{  t}} 
=\dfrac { t_0} {\sqrt {1- \dfrac {v^2} {c^2}}} =  t
\thinspace ,
\label{eq:qt_measured_int_define}
\end{equation} 
which is the Lorentz time dilation generally accepted as a verified experimental result.

Next, let us consider the same experiment in the observer's reference frame. Clearly, we expect to obtain the same experimental result even though the platform is now moving away from the observer with a constant velocity $-\vec v$. The starting time of the measurement and the clock synchronization time is zero, as in the source reference frame. However, the end time-point is now given by the conjugate quaternion due to space inversion when switching from the source to the observer reference frame, 

\begin{equation}
\boldsymbol{\bar t }
=\left ( \thinspace t_0 \thinspace, -\thinspace  \dfrac  {\vec v} {c}  \thinspace t\thinspace \right )
\thinspace ,
\label{eq:qt_conjugate_time_interval}
\end{equation}
as shown in  Fig.~\ref{fig:QT_Polar_Frame_Observer}.

The measured time-interval duration in the observer reference frame is given by,
\begin{equation}
|\boldsymbol{{ \bar  t }}| 
= \sqrt{\boldsymbol{ t}\thinspace \boldsymbol{{  \bar  t}}}
=\sqrt{\boldsymbol{{\bar   t}}\thinspace  \boldsymbol{  t}} 
=\dfrac { t_0} {\sqrt {1- \dfrac {v^2} {c^2}}} =  t
\thinspace .
\label{eq:qt_conj_int_define}
\end{equation} 
As expected, the measured time duration remains the same despite the conjugate form of the time interval in the observer reference frame.

\begin{figure}
	\includegraphics{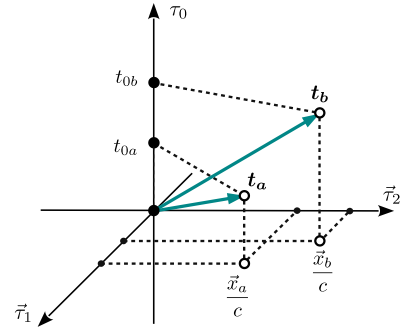}
	\caption{\label{fig:QT_Source_2_Observers} A three-dimensional representation of two arbitrary time points.}
\end{figure}

Next, let us define an arbitrary quaternion time interval as a difference between any two quaternion time points, $\boldsymbol {  t_a}$ and $\boldsymbol {  t_b}$, as can be seen in Fig.~\ref{fig:QT_Source_2_Observers},

\begin{equation}
\begin{cases}
\boldsymbol{  t_a} = \left ( t_{0a} \thinspace,   \dfrac {\vec{ x_a}}  {c}  \thinspace \right) ,
\\
\\
\boldsymbol{  t_b} = \left ( t_{0b},  \dfrac {\vec{ x_b}}  {c}  \thinspace \right).
\end{cases}		
\label{eq:qt_source_points}
\end{equation}
Let us calculate the time interval at the  observer location,
\begin{equation}
\boldsymbol{     t} 
=\boldsymbol{   t_b}  - \boldsymbol{  t_a} 
=  \left ( t_{0}, \dfrac {\vec{ x}}  {c} \right ) 
=  \left ( t_{0}, \dfrac {\vec{ v}}  {c} \thinspace t\right ), 		
\label{eq:qt_source_interval}
\end{equation}
where we define the velocity with arbitrary direction as, $\vec v = \vec x /t$.
Similarly, in the observer frame,
\begin{equation}
\boldsymbol{   \bar  t} 
=\boldsymbol{  \bar t_b}  - \boldsymbol{ \bar t_a} 
=  \left ( t_{0},- \dfrac {\vec{ x}}  {c} \right ) 
=  \left ( t_{0}, -\dfrac {\vec{ v}}  {c} \thinspace t\right ) 	.	
\label{eq:qt_observer_interval}
\end{equation}

It appears therefore that quaternions are capable of describing time transformation between the signal source and observer for arbitrary relative motion between them, including propagation and rotation. 

\begin{subfigures}
	\begin{figure}
		\includegraphics{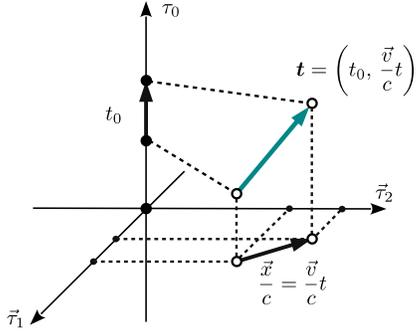}
		\caption{\label{fig:QT_Source_Moving_Observer} A simplified three-dimensional representation of the arbitrary quaternion time interval in the source reference frame.}
	\end{figure}
	\begin{figure}
		\includegraphics{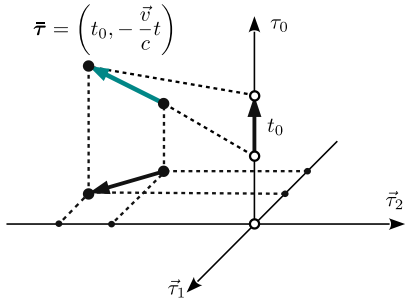}
		\caption{\label{fig:QT_Observer_Moving_Source} A simplified three-dimensional representation of the quaternion time interval in the observer reference frame.}
	\end{figure}
\end{subfigures}

In  Fig.~\ref{fig:QT_Source_Moving_Observer} and Fig.~\ref {fig:QT_Observer_Moving_Source}, we demonstrate an arbitrary quaternion time interval in the source and observer reference frames, describing arbitrary relative motion composed of propagation and rotation are possible.

\begin{subfigures}
	\begin{figure}
		\includegraphics{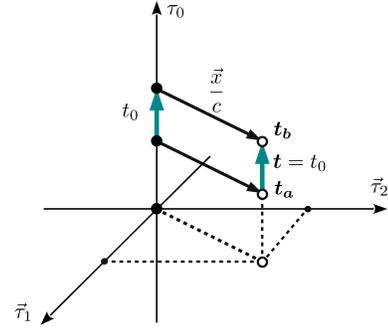}
		\caption{\label{fig:stationary_clock_interval} A three-dimensional representation of the stationary observer in the source reference frame.}
	\end{figure}
	\begin{figure}
		\includegraphics{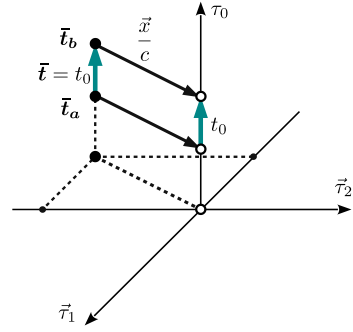}
		\caption{\label{fig:stationary_observer_interval} A three-dimensional representation of the stationary clock in the observer reference frame.}
	\end{figure}
\end{subfigures}

Let us consider a measurement of the time interval displayed on the zero-point clock in the source frame, $  t_0$, by a stationary observer located at a space location $\vec x$ away from the clock.  The initial and final time quaternions are
\begin{equation}
\begin{cases}
\boldsymbol{  t_a} = \left ( t_{0a},  \dfrac {\vec{ x}}  {c} \right) 
\\
\\
\boldsymbol{  t_b} =  \left ( t_{0b},  \dfrac {\vec{ x}}  {c} \right)  
\end{cases}		
\label{eq:q_st_clock_interval}
\end{equation}
Let us calculate the time interval at the  observer location,
\begin{equation}
\boldsymbol{     t} 
=\boldsymbol{   t_b}  - \boldsymbol{  t_a} 
=  t_{0b} -  t_{0a} =   t_0,		
\label{eq:q_st_interval_difference}
\end{equation}

Similarly, in the observer frame, we obtain,
\begin{equation}
\begin{cases}
\boldsymbol{\bar {  t}_a}=  \left ( t_{0a},  -\dfrac {\vec{ x}}  {c} \right)   
\\
\\
\boldsymbol{\bar {  t}_b} = \boldsymbol{  t_b} =  \left ( t_{0b}, - \dfrac {\vec{ x}}  {c} \right) 
\end{cases}		
\end{equation}
We can now calculate the measured time interval

\label{eq:q_st_observer_interval}
\begin{equation}
{\boldsymbol{   \bar{  t}}} = \boldsymbol{   \bar{  t}_b} - \boldsymbol{   \bar{  t}_a}
=  t_{0b}- t_{0a} 
=  t_0
\thinspace .		
\label{eq:q_conj_st_interval_difference}
\end{equation}
Therefore for the stationary source and observer, measurement of the time interval results in the same value, $t_0$. This is due to the fact that  time-of-flight of photons cancels out.  

Therefore, we conclude that quaternion time intervals,

\begin{equation}
\begin{cases}
\boldsymbol{  t } 
=    \left ({\thinspace  { t_0}}   \thinspace ,  \dfrac {\vec v}  {c} \thinspace t  \thinspace \right )  
=\dfrac {t_0}    {\sqrt {1- \dfrac {v^2} {c^2}}}
\thinspace \exp \left ( {  \vec\imath \thinspace \theta}  \right )  
\thinspace ,
\\
\\
\boldsymbol{ \bar t } 
=    \left ({\thinspace  { t_0}}   \thinspace ,  -\dfrac {\vec v}  {c} \thinspace t  \thinspace \right ) 
=  \dfrac {   t_0 } 
{\sqrt {1- \dfrac {v^2} {c^2}}} \thinspace \exp \left ( { - \vec\imath \thinspace \theta}  \right ) \thinspace, 
\end{cases}	
\label{eq:qt_time_combined}
\end{equation}
describe the measured time interval at the signal source, $t_0$, as well as the measured interval at the observer location, $t$, during relative motion between the source and observer, expressed by the normalized vector velocity, $\vec v /c$. The regular quaternion, $\boldsymbol t$, describes time in the source reference frame while the quaternion conjugate, $\boldsymbol{ \bar t} $, describes time in the observer reference frame. This physical interpretation  is similar to the relativistic Doppler effect approach \cite{Moriconi}, however using quaternion mathematical formalism. Note that the physical interpretation of the quaternion time intervals includes description of experimental measurements as an integral part of the theory.

\section{Inverse Time Interval - Quaternion Energy, Mass, and  Matter} %

Let us consider a periodic signal source such as a clock on a kitchen wall or a distant quasar. Assume that such a signal source produces a wave signal with a period described by a time interval, $t_0$. Then, assume that the periodic signal in the observer reference frame can be described by a wave with a period,  ${ t}$. Let us introduce the quaternion definition of frequency, $\boldsymbol {  \omega}$, by using the quaternion inverse of the time interval,

\begin{equation}
\begin{cases}
\boldsymbol { \bar \omega} =
\dfrac {2\pi  }  {{\boldsymbol    t} }
=   
\dfrac {{2\pi \boldsymbol{ \bar  t}}}  {   t^2 }
= 
\omega
\left ( \dfrac {  t_0} {   t} \thinspace , -
\dfrac {\vec v} { c} \right ),
\\
\\
\boldsymbol {  \omega} =
\dfrac {2\pi  }  {{\boldsymbol{   \bar  t}} }
=  
\dfrac {2\pi \boldsymbol{  t}}  {   t^2 }
= 
\omega
\left ( \dfrac {  t_0} {   t} \thinspace , \thinspace 
\dfrac {\vec v} { c} \right ),
\end{cases}
\label{eq:q_freq_define}
\end{equation}
where we define the absolute value of frequency, 
\begin{equation}
\omega 
=|\boldsymbol {   \omega}|
=|\boldsymbol {  \bar \omega} |
=\dfrac {2\pi  }  {{   t} }
\thinspace.
\label{eq:zero_point_frequency}
\end{equation}   
Note that the above definition is possible due to well defined quaternion multiplicative inverse. Note that in the observer reference frame, the space-time interval is given by a complex-conjugate quaternion, $\boldsymbol \bar t$, resulting in the frequency with a positive vector component, $\vec \omega = (\vec v /c ) t$, due to mathematical properties of quaternion division. On the other hand, in the source frame of reference, where space-time is represented by the regular quaternion, $\boldsymbol  t$, the frequency assumes quaternion conjugate form with a negative vector part, $\vec \omega = -(\vec v /c ) t$. Since we are primarily interested in physical results in the observer reference frame, we assume that quaternion space-time is given by the conjugate quaternion, $\boldsymbol {\bar  t}$, while the frequency is given by the regular quaternion, $\boldsymbol \omega$.  

Defining the zero-point frequency component from (\ref{eq:q_freq_define}), 
\begin{equation}
{  \omega_0} 
=  \omega   \dfrac {  t_0} {   t}
=\omega  \thinspace {\sqrt {1-\dfrac {v^2} {c^2} }}
\thinspace,
\label{eq:qt_new_doppler}
\end{equation}  
we can re-write frequency in the observer frame of reference as,
\begin{equation}
\boldsymbol {   \omega}
=  
\left ( \omega_0 \thinspace ,
\thinspace   \dfrac {\vec v} {c }  \thinspace\omega \thinspace  \right ),  
\label{eq:q_freq_simp}
\end{equation} 

Let us assume that we can express energy in the observer reference frame using frequency (\ref{eq:q_freq_simp}),

\begin{equation}
\boldsymbol {   \epsilon }
= \hbar \thinspace\boldsymbol {   \omega} =
\left ( {   \epsilon_0} \thinspace ,  \thinspace \dfrac {\vec{v}} { c }\thinspace   \epsilon \right ).
\label{eq:qt_energy_define}
\end{equation}

Next, we apply the same procedure to define mass from energy using (\ref{eq:qt_energy_define}),
\begin{equation}
\boldsymbol {  m}
= \dfrac {\thinspace\boldsymbol {    \epsilon}} {c^2} 
= \left ( {  m_0} \thinspace ,  \thinspace \dfrac {  \thinspace\vec{v}} { c }\thinspace  m  \thinspace\right )
= \left ( {  m_0}\thinspace , \thinspace \dfrac { \vec{p}} { c }\thinspace\right ),
\label{eq:qt_energy_mass}
\end{equation}  
where we defined the momentum as, $\vec p=m \vec v$, and obtained the mass transformation,

\begin{equation}
m
=   \dfrac  {m_0 }
{\sqrt {1-\dfrac {v^2} {c^2} }}
\thinspace.
\label{eq:qt_mass}
\end{equation}  
This leads to the traditional energy-momentum relation,

\begin{equation}
\epsilon^2 =   m_0^2 c^4+  p^2 c^2.
\label{eq:qt_disperssion_energy}
\end{equation}

For large velocities, $v \simeq c$ ,

\begin{equation}
\begin{cases}
m_0 
=   m \thinspace 
\sqrt {1-\dfrac {v^2} {c^2} }
\simeq 0
\thinspace ,
\\
\\
\dfrac {\vec v} {c} 
\simeq \dfrac {\vec c} {c}
\simeq \vec \imath
\thinspace .	
\end{cases}		
\label{eq:qt_high_speed_mass}
\end{equation}

This results in an  approximation for mass of particles with a zero rest-mass, such as photons,

\begin{equation}
\boldsymbol {  m} 
\simeq \left (0\thinspace, \thinspace \vec \imath \thinspace   m \right )
= \left (0 \thinspace, \thinspace \vec \imath \thinspace \thinspace \dfrac {\hbar \thinspace{    \omega}} {c^2}  \thinspace\right ).
\label{eq:qt_photon_mass}
\end{equation}  

For small velocities, $v \ll c$ , and $\vec p \sim 0$,  
\begin{equation}
\boldsymbol {   m} \simeq   m_0 
\thinspace.
\label{eq:slow_mass}
\end{equation}

Next, let us introduce quaternion density of mass in the observer frame,
\begin{equation}
\boldsymbol {\rho}
=  \left ( \rho_{0m}, \thinspace \dfrac {\vec{v}}   {c} \thinspace {\rho_m} \thinspace \right )
= \left ( \rho_0, \thinspace \dfrac {\vec{j}_m} {c} \thinspace \right ).
\label{eq:qt_em_charge_density}
\end{equation}

Similarly, assume that we can propose quaternion definitions of charge,
\begin{equation}
\boldsymbol{ q} 
=  \left ( q_0 ,  \dfrac {\vec{v} \thinspace  } {c} \thinspace {q} \thinspace \right ),
\label{eq:q_charge}
\end{equation}  

which leads to the quaternion density of charge in the observer frame,
\begin{equation}
\boldsymbol {\rho} 
=  \left ( \rho_0, \thinspace \dfrac {\vec{v}}   {c} \thinspace {\rho_q} \thinspace \right )
= \left ( \rho_{0q}, \thinspace \dfrac {\vec{j}_q} {c} \thinspace \right )
\label{eq:em_energy_em}
\end{equation}

Also, let us assume that any interaction can be described by an interaction potential function given in the observer reference frame as a quaternion,

\begin{equation}
{\boldsymbol{\phi}} = ( \phi_0 \thinspace,\thinspace\vec{\phi}_v\thinspace)
=\left ( \phi_0 \thinspace,\thinspace\dfrac {\vec v} {c}{\phi}\thinspace \right),
\label{eq:qt_potential}
\end{equation}
where, $\phi_0$, is the static potential at the signal source, while, ${\vec \phi}_v = \vec v /c \phi$, is the vector potential due to the motion of the source relative to the observer. As usual, we can define the potential measured by the observer as the quaternion norm,
\begin{equation}
\phi = |\boldsymbol{{ \phi }}| 
= |\boldsymbol{{ \bar \phi }}| 
=\dfrac { \phi_0} {\sqrt {1- \dfrac {v^2} {c^2}}} 
\thinspace .
\label{eq:qt_phi_norm}
\end{equation} 

It appears that any physical property under observation consists of the static zero-point component corresponding to this property measured at rest. In addition, there is an imaginary vector correction term proportional to, $\vec v/ c$, and accounting for relative motion between the source and observer.

\section{Quaternion Potential Derivative and Generic Fields} %

We take advantage of quaternion division in order to deduce a proper quaternion differential operator in the observer reference frames. Also, we assume that we can assign the physical meaning of the generalized force fields to the quaternion derivative of the  potential function.

Using the definition of the quaternion multiplicative inverse,

\begin{equation}
\begin{cases}
\boldsymbol{ t^{-1}}
=\dfrac {\boldsymbol{\bar t}}  { t^2} ,
\\
\\
\boldsymbol{\bar t^{-1}}
=\dfrac {\boldsymbol{ t}}  { t^2} ,
\end{cases}
\label{eq:qt_inverse}
\end{equation}
we define the quaternion differential operators,

\begin{equation}
\begin{cases}
{\boldsymbol{\bar \nabla}}
=\dfrac {1} {c}  \thinspace \dfrac {d} {d\boldsymbol {  t}}
= \left (\dfrac {\partial } {c \partial t_0},
-\vec{\imath}_1 \dfrac {\partial } { \partial x_1 },
-\vec{\imath}_2 \dfrac {\partial } { \partial x_2 },
-\vec{\imath}_3 \dfrac {\partial } {\partial x_3 } \right ),
\\
\\
{\boldsymbol{\nabla}} 
=\dfrac {1} {c} \thinspace \dfrac {d} { d \boldsymbol { \bar  t}}
= \left (\dfrac {\partial } {c \partial t_0}, 
+\vec{\imath}_1 \dfrac {\partial } { \partial x_1 },
+\vec{\imath}_2 \dfrac {\partial } { \partial x_2 },
+\vec{\imath}_3 \dfrac {\partial } {\partial x_3 } \right ) .
\end{cases}
\label{eq:qt_del}
\end{equation}
We write the four-dimensional gradients in the simplified vector notation as,
\begin{equation}
\begin{cases}
{\boldsymbol{\bar \nabla}}
= \left(\dfrac {\partial } {c \partial t_0} \thinspace, - \vec{\nabla}\right)
= \left({\nabla_0}\thinspace, - \vec{\nabla}\right),
\\
\\
{\boldsymbol{\nabla}} 
= \left (\dfrac {\partial } {c \partial t_0}\thinspace, \thinspace \thinspace \vec{\nabla}\right)
= \left ({\nabla_0}\thinspace, \thinspace \thinspace \vec{\nabla}\right).
\end{cases}
\label{eq:qt_del_simple}
\end{equation}
Thus, the correct form of the quaternion differential operator assumes the conjugate form, $ \boldsymbol{\bar \nabla}$, in the source reference frame.
On the other hand, in the observer reference frame, the expression for the differential operator has the regular quaternion form , $ \boldsymbol{\nabla}$, due to space inversion during the division operation in (\ref{eq:qt_inverse}).

Since we are primarily interested in the reference frame of the measuring apparatus, which is the observer reference frame, we will use the form of the derivative operator given by $ \boldsymbol{\nabla}$.

Then, we use the definition of quaternion multiplication for any two quaternions, $\boldsymbol a$ and $\boldsymbol b$, 

\begin{equation}
\begin{cases}
\thinspace \thinspace \boldsymbol{a} \thinspace \boldsymbol{b}  
= \left ( \thinspace a_0 b_0 - \vec{a} \cdot \vec {b}, \thinspace 
a_0\vec{b} + b_0 \vec {a} +\vec{a}\times \vec{b}\thinspace \thinspace \right),
\\

\thinspace \thinspace \boldsymbol{b} \thinspace \boldsymbol{a}  
=  \left (\thinspace  a_0 b_0 - \vec{a} \cdot \vec {b}, 
\thinspace a_0\vec{b} + b_0 \vec {a} ,
-\vec{a}\times \vec{b}\thinspace \thinspace \right) \thinspace ,
\end {cases}
\label{eq:q_mult}
\end{equation}
to define two force fields as derivatives of the potential function,
\begin{equation}
\begin{cases}
\boldsymbol{\mathcal{F}^+}
=-\boldsymbol{{\phi}} \thinspace {\boldsymbol{\nabla}  },
\\
\\
\boldsymbol{\mathcal{F}^-}
= -{\boldsymbol{\nabla} }  {\boldsymbol{{\phi}}}  .
\end {cases}
\label{eq:qt_force define}
\end{equation}

Note that the two derivatives are due to non-commutativity of the quaternion multiplication resulting in the right and left derivatives. 

Then applying (\ref{eq:qt_potential}) and (\ref{eq:q_mult}) to (\ref{eq:qt_force define}), we obtain general expressions for the quaternion force fields,

\begin{equation}
\begin{cases}
\boldsymbol{\mathcal{F}^+}
=  \left(-{\nabla_0\phi_0} +\vec {\nabla} \cdot \vec{\phi}_v  , \thinspace
-{\nabla_0}\vec{\phi}_v  -\vec{\nabla}\phi_0 
+\vec{\nabla} \times \vec{\phi}_v \thinspace\right),
\\
\\
\boldsymbol{\mathcal{F}^-}
=  \left(-{\nabla_0\phi_0} +\vec {\nabla} \cdot \vec{\phi}_v  , \thinspace
-{\nabla_0}\vec{\phi}_v  -\vec{\nabla}\phi_0  
-\vec{\nabla} \times \vec{\phi}_v \thinspace\right).
\end {cases}
\label{eq:qt_general_Lorentz_force}
\end{equation}
We can consider (\ref{eq:qt_force define}) and (\ref{eq:qt_general_Lorentz_force}) as the generalized Lorentz field expressions.

Let us look for single-valued functions defining the field components by using commutator and anti-commutator relations,

\begin{equation}
\begin{cases}
\boldsymbol{\mathcal{F}_a}
= \dfrac  {1} {2} \left ( \boldsymbol{\mathcal{F}\thinspace^+} + \boldsymbol{\mathcal{F}\thinspace^-}  \right),
\\
\\
\boldsymbol{\mathcal{F}_c}
= \dfrac  {1} {2} \left ( \boldsymbol{\mathcal{F}\thinspace^+} - \boldsymbol{\mathcal{F}\thinspace^-}  \right)
.
\end {cases}
\label{eq:qt_fileds_define}
\end{equation}

This results in two types of the generalized field components that can be further expanded from (\ref{eq:qt_general_Lorentz_force}) and (\ref{eq:qt_fileds_define}), 

\begin{equation}
\begin{cases}
\boldsymbol{\mathcal{F}_a}
=  \left({\mathcal{F}_0}\thinspace, \thinspace
\vec{\mathcal{F}}_c  
\thinspace\right)
=  \left(-\dfrac {\partial \phi_0} {c \partial t_0} +\vec {\nabla} \cdot \vec{\phi}_v  \thinspace, \thinspace
- \dfrac {\partial \vec{\phi}_v} {c \partial t_0}  -\vec{\nabla}\phi_0  
\thinspace\right),
\\
\\
\boldsymbol{\mathcal{F}_c}
=\vec{\mathcal{F}}_c
=  
\vec{\nabla} \times \vec{\phi}_v \thinspace.
\end {cases}
\label{eq:qt_unified_fields}
\end{equation}

Using the velocity dependent vector potential $\phi_v = (\vec v /c ) \phi$, we can express the field components in the velocity dependent form,

\begin{equation}
\begin{cases}
{\mathcal{F}_0}
=  -\dfrac {\partial \phi_0} {c \partial t_0} +\phi \vec {\nabla} \cdot \dfrac {\vec{v}} {c} + (\vec {\nabla}\phi)  \cdot \dfrac {\vec{v}} {c}\thinspace,
\\
\\
\vec{\mathcal{F}_a}
=   \thinspace
- \dfrac {\partial (\vec{v}\phi)} {c^2 \partial t_0}  -\vec{\nabla}\phi_0  ,
\\
\\
\vec{\mathcal{F}}_c
=  
\phi \left (\vec {\nabla} \times \dfrac {\vec{v}} {c}\right ) +(\vec {\nabla}\phi)  \times \dfrac {\vec{v}} {c}\thinspace.
\end {cases}
\label{eq:qt_unified_velocity_fields}
\end{equation}

The total fields can be now written in terms of the field components using (\ref{eq:qt_fileds_define}),
\begin{equation}
\begin{cases}
\boldsymbol{\mathcal{F}^+}
=\boldsymbol{\mathcal{F}_a}+\boldsymbol{\mathcal{F}_c}
\\
\\
\boldsymbol{\mathcal{F}^-}
=\boldsymbol{\mathcal{F}_a}-\boldsymbol{\mathcal{F}_c} \thinspace  .
\end {cases}
\label{eq:qt_force total}
\end{equation}

For the stationary case, when, $\vec v\sim 0$, the field components become,
\begin{equation}
\begin{cases}
\boldsymbol{\mathcal{F}^+} 
=\boldsymbol{\mathcal{F}^-}
=\boldsymbol{\mathcal{F}_a}
\simeq \left(-\dfrac {\partial \phi_0} {c \partial t_0} \thinspace, \thinspace
 -\vec{\nabla}\phi_0  
\thinspace\right)
\\
\\
\boldsymbol{\mathcal{F}_c}
=\vec{\mathcal{F}}_c
\simeq \vec 0 \thinspace,
\end {cases}
\label{eq:qt_unified_stationary_fields}
\end{equation}

which further reduce to the classical field expression for the stationary static field, when $\partial \phi_0 / \partial t_0 \sim 0$,

\begin{equation}
\begin{cases}
\boldsymbol{\mathcal{F}^+} 
=\boldsymbol{\mathcal{F}^-}
=\vec{\mathcal{F}}_a
\simeq \thinspace 
-\vec{\nabla}\phi_0 \thinspace . 

\\
\\
\boldsymbol{\mathcal{F}_c}
\simeq\vec{\mathcal{F}}_c
=  \vec 0 \thinspace.
\end {cases}
\label{eq:qt_classical_fields}
\end{equation}

Thus, we obtained generalized field equations for an arbitrary physical interaction defined by a quaternion potential function, $\boldsymbol \phi$. One of the field components, $\boldsymbol{\mathcal{F}_a}$, is a full quaternion, with both scalar and vector parts, which describe time-dependent fluctuations and linear propagation. The other field component, $\boldsymbol{\mathcal{F}_c}$, is a three-dimensional pure vector, describing a torsion field and rotation in the presence of motion. The field components combine into two expressions for the total quaternion field, $\boldsymbol{\mathcal{F}\thinspace^+}$ and $\boldsymbol{\mathcal{F}\thinspace^-}$, which reduce to the classical field expression for the static stationary case. Note that the new field expressions were derived from a velocity dependent quaternion potential and consequently depend on the velocity of the body under investigation relative to the observer measuring apparatus.

\section{Second Derivative and Generalized Density of Matter} %

Next, we introduce the second derivative of the potential function. Also, let us assume that the physical interpretation of the second derivative of the potential function, $\boldsymbol \phi$, corresponds to the generalized density of matter, $ \boldsymbol \rho$, similar to Poisson's equation.

Assume generalized density of matter in the observer reference frame,
\begin{equation}
\boldsymbol {\rho}
=  \left ( \rho_0, \thinspace \dfrac {\vec{v}}   {c} \thinspace {\rho} \thinspace \right )
= \left ( \rho_{0}, \thinspace \dfrac {\vec{j}} {c} \thinspace \right ).
\label{eq:em_generic_density}
\end{equation}
We can define derivatives by applying left and right differentiation  to the two kinds of the generalized Lorentz fields, $\boldsymbol{\mathcal{F}^+}$ and  $\boldsymbol{\mathcal{F}^-}$, using (\ref{eq:qt_force define}),
\begin{equation}
\begin{cases}
 {\boldsymbol {\rho^+}}
=\boldsymbol{\mathcal{F}^+} {\boldsymbol{\nabla}}
=-\boldsymbol{{\phi}}  {\boldsymbol{\nabla}}  {\boldsymbol{\nabla}},
\\
\\
 {\boldsymbol {\rho^-}}
={\boldsymbol{\nabla} }\boldsymbol{\mathcal{F}^-}
=-{\boldsymbol{\nabla} }{\boldsymbol{\nabla} }  {\boldsymbol{{\phi}}} ,
\\
\\
{\boldsymbol {\rho^\sim}}
= {\boldsymbol{\nabla} \boldsymbol{\mathcal{F}^+}}
= - {\boldsymbol{\nabla}}   \boldsymbol{{\phi}} {\boldsymbol{\nabla}},
\\
\\
{\boldsymbol {\rho^\sim}}
= \boldsymbol{\mathcal{F}^-} {\boldsymbol{\nabla}}
= - {\boldsymbol{\nabla}}   \boldsymbol{{\phi}} {\boldsymbol{\nabla}}.
\end {cases}
\label{eq:qt_second_der_density}
\end{equation}

We obtain four types of matter density, however, the last two are the same because quaternion multiplication is distributive,

\begin{equation}
({\boldsymbol{\nabla}}   \boldsymbol{{\phi}} ) {\boldsymbol{\nabla}} 
={\boldsymbol{\nabla}}   (\boldsymbol{{\phi}} {\boldsymbol{\nabla}})
={\boldsymbol{\nabla}}   \boldsymbol{{\phi}} {\boldsymbol{\nabla}}.
\label{eq:qt_distributive_mult}
\end{equation}

From (\ref{eq:qt_force total}) and (\ref{eq:qt_second_der_density}),
\begin{equation}
\begin{cases}
 {\boldsymbol {\rho^+}}
= \boldsymbol{\mathcal{F}_a}{\boldsymbol{\nabla} }+\boldsymbol{\mathcal{F}_c}{\boldsymbol{\nabla} },

\\
\\
 {\boldsymbol {\rho^-}}
={\boldsymbol{\nabla} } \boldsymbol{\mathcal{F}_a}-{\boldsymbol{\nabla} }\boldsymbol{\mathcal{F}_c},

\\
\\

 {\boldsymbol {\rho^\sim}}
= {\boldsymbol{\nabla} }\boldsymbol{\mathcal{F}_a}+{\boldsymbol{\nabla} }\boldsymbol{\mathcal{F}_c}
= \boldsymbol{\mathcal{F}_a}{\boldsymbol{\nabla} }-\boldsymbol{\mathcal{F}_c}{\boldsymbol{\nabla} }.

\end {cases}
\label{eq:qt_second_der_expand}
\end{equation}

Calculating left and right derivatives of the field components using (\ref{eq:qt_unified_fields}),
\begin{equation}
\begin{cases}
{{\boldsymbol {\nabla}}}\thinspace  {{\boldsymbol {\mathcal{F}_a}}}
=  \left ({\nabla_0}\mathcal{F}_0
-\vec{\nabla}\cdot\vec{\mathcal{F}}_a 
\thinspace \thinspace , \thinspace 
{\nabla_0}\vec{\mathcal{F}}_a
+\vec{\nabla} \mathcal{F}_0
+\vec{\nabla}\times \vec{\mathcal{F}}_a \right )
\\
\\
{{\boldsymbol {\mathcal{F}_a}}} \thinspace  { \boldsymbol {\nabla}} 
=  \left ({\nabla_0}\mathcal{F}_0
-\vec{\nabla}\cdot\vec{\mathcal{F}}_a 
\thinspace \thinspace , \thinspace 
{\nabla_0}\vec{\mathcal{F}}_a
+\vec{\nabla} \mathcal{F}_0
-\vec{\nabla}\times \vec{\mathcal{F}}_a \right ).
\end {cases}
\label{eq:qt_second_translational}
\end{equation}
Similarly, for the torsion field,

\begin{equation}
\begin{cases}
{\boldsymbol {\nabla}}\thinspace  {{\boldsymbol {\mathcal{F}_c}}}
=\left (\vec{\nabla}\cdot\vec{\mathcal{F}_c} \thinspace \thinspace,\thinspace {\nabla_0}\vec{\mathcal{F}}_c
+\vec{\nabla}\times \vec{\mathcal{F}_c} \right )
\\
\\
{{\boldsymbol {\mathcal{F}_c}}}\thinspace {\boldsymbol {\nabla}} 
=\left (\vec{\nabla}\cdot\vec{\mathcal{F}_c} \thinspace \thinspace,\thinspace {\nabla_0}\vec{\mathcal{F}}_c
-\vec{\nabla}\times \vec{\mathcal{F}_c} \right )
\end {cases}
\label{eq:qt_second_tortionl}
\end{equation}

Since divergence of curl is zero,

\begin{equation}
\vec{\nabla}\cdot\vec{\mathcal{F}_c} 
=  \vec{\nabla}\cdot \left(\vec{\nabla}\times \vec{\phi}_v \thinspace \right)= 0,
\label{eq:qt_divergence_of_rot}
\end{equation}
we obtain for the torsion field,

\begin{equation}
\begin{cases}
{\boldsymbol {\nabla}}\thinspace  {{\boldsymbol {\mathcal{F}_c}}}
=\left (0 \thinspace \thinspace,\thinspace {\nabla_0}\vec{\mathcal{F}}_c
+\vec{\nabla}\times \vec{\mathcal{F}_c} \right )
\\
\\
{{\boldsymbol {\mathcal{F}_c}}}\thinspace {\boldsymbol {\nabla}} 
= \left (0\thinspace \thinspace,\thinspace {\nabla_0}\vec{\mathcal{F}}_c
-\vec{\nabla}\times \vec{\mathcal{F}_c} \right ).
\end {cases}
\label{eq:qt_second_torsion_simple}
\end{equation}

From the last equation in (\ref{eq:qt_second_der_expand}), we derive an equality,
\begin{equation}
\dfrac {\partial \vec{\mathcal{F}}_c} {c \partial t_0} 
+
\vec{\nabla}\times \vec{\mathcal{F}}_a = 0.
\label{eq:qt_equal_densities}
\end{equation}

Writing quaternion density  in terms of scalar density and vector current density components,

\begin{equation}
\begin{cases}

{ \boldsymbol \rho^+}  
=  \left ( \rho_0, \dfrac { \vec j^+} {c} \thinspace\right ),
\\
\\

{ \boldsymbol \rho^-}   
=  \left ( \rho_0, \dfrac {\vec j^- } {c}\thinspace \right ),

\\
\\
 \boldsymbol \rho^\sim  
=  \left ( \rho_0, \dfrac {{\vec j}^\sim } {c} \thinspace\right ),
\end {cases}
\label{eq:qt_density_components}
\end{equation}

we obtain complete set of generic equations for the three distinct types of the quaternion density of matter,
\begin{equation}
\begin{cases}

 { \rho_0}  
= -\dfrac {\partial \mathcal{F}_0} {c \partial t_0} 
+\vec{\nabla}\cdot\vec{\mathcal{F}}_a \thinspace , 

\\
\\

 \dfrac {\vec j^+} {c} 
= 
-\dfrac {\partial \vec{\mathcal{F}}_a} {c \partial t_0} 
-\vec{\nabla} \mathcal{F}_0 
\thinspace 
-\vec{\nabla}\times \vec{\mathcal{F}_c}
+\thinspace 2 \thinspace
\dfrac {\partial \vec{\mathcal{F}}_c} {c \partial t_0},
\\
\\

 \dfrac {\vec j^-}  {c}
= 
-\dfrac {\partial \vec{\mathcal{F}}_a} {c \partial t_0} 
-\vec{\nabla} \mathcal{F}_0 
-\vec{\nabla}\times \vec{\mathcal{F}_c}
-\thinspace 2\thinspace
\dfrac {\partial \vec{\mathcal{F}}_c} {c \partial t_0} ,
\\
\\

 \dfrac {\vec j^\sim} {c} 
=  
-\dfrac {\partial \vec{\mathcal{F}}_a} {c \partial t_0} 
-\vec{\nabla} \mathcal{F}_0 
+\vec{\nabla}\times \vec{\mathcal{F}_c},
\\
\\
 \thinspace 
\dfrac {\partial \vec{\mathcal{F}}_c} {c \partial t_0} 
+
\vec{\nabla}\times \vec{\mathcal{F}}_a = 0,
\\
\\
\vec{\nabla}\cdot\vec{\mathcal{F}_c} .
= 0
\end {cases}
\label{eq:qt_generic_matter_densities}
\end{equation}

The different current components result from the left and right differentiation of two different charges in the generalized Lorentz field (\ref{eq:qt_general_Lorentz_force}). Importantly, the matter density equations are derived directly from the Lorentz field expressions by differentiation, thus avoiding incompatibility between them, which appears between the traditional Lorentz force and Maxwell equations.

\section{Quaternion Electromagnetic Fields and Charge Density} %

Let us consider electromagnetic interaction expressed by a quaternion electromagnetic potential, $\boldsymbol \phi$, in the observer reference frame, where we introduce the electric and magnetic fields using (\ref{eq:qt_fileds_define}),

\begin{equation}
\begin{cases}
\boldsymbol{\mathcal{F}_a}
= \boldsymbol{\mathcal E}
=  \left(\thinspace \mathcal{E}_0 \thinspace, \thinspace\thinspace
\mathcal{\vec E}\thinspace\right),
\\
\\
\boldsymbol{\mathcal{F}_c}
= \boldsymbol{\mathcal B}
=  \left(\thinspace 0, \thinspace\thinspace
 \mathcal{\vec B}\thinspace\right).
\end {cases}
\label{eq:qt_em_define}
\end{equation}

As we can see, the electric field is a full quaternion, with both the scalar and vector components. On the other hand the magnetic field is purely a vector field. We derive full expressions for the electromagnetic fields from (\ref{eq:qt_unified_fields}) and (\ref{eq:qt_em_define}), 
\begin{equation}
\begin{cases}
\mathcal{E}_0
= \thinspace -\dfrac {\partial \phi_0} {c \partial t_0}  + \vec {\nabla} \cdot \vec{\phi}_v ,
\\
\\	
\mathcal{\vec E}
=  - \dfrac {\partial \vec{\phi}_v} {c \partial t_0}- \vec{\nabla} \phi_0 ,
\\
\\
\mathcal{\vec B}
= \thinspace\vec{\nabla}\times \vec{\phi}_v\thinspace ,
\end{cases} 
\label{eq:qt_em_fields}
\end{equation}
which reminds of the traditional expressions for the electric and magnetic fields, with the exception of 
of a scalar component of the electric field, $\mathcal{E}_0$, which is not present in the traditional approach. 

Using the velocity dependent vector potential, $\phi_v = (\vec v /c ) \phi$, we obtain from (\ref{eq:qt_unified_velocity_fields}) and (\ref{eq:qt_em_fields}),

\begin{equation}
\begin{cases}
\mathcal{E}_0
= -\dfrac {\partial \phi_0} {c \partial t_0} +\phi \thinspace \vec {\nabla} \cdot \dfrac {\vec{v}} {c} + (\vec {\nabla}\phi)  \cdot \dfrac {\vec{v}} {c}\thinspace,
\\
\\	
\mathcal{\vec E}
=\thinspace
- \dfrac {\partial (\vec{v}\phi)} {c^2 \partial t_0}  -\vec{\nabla}\phi_0  
\thinspace,
\\
\\
\mathcal{\vec B}
=  
\phi \left (\vec {\nabla} \times \dfrac {\vec{v}} {c}\right ) +(\vec {\nabla}\phi)  \times \dfrac {\vec{v}} {c}\thinspace.
\end{cases} 
\label{eq:qt_em_velocity_fields}
\end{equation}

By applying (\ref{eq:qt_em_define}) to the definition of the generalized force fields (\ref{eq:qt_force total}), we obtain two quaternion expressions for the total electromagnetic fields, 
\begin{equation}
\begin{cases}
\boldsymbol{\mathcal{F}}^+
=  \left(\mathcal{E}_0\thinspace, \thinspace\thinspace
\mathcal{\vec E} 
+\mathcal{\vec B}\thinspace\right),

\\
\\
\boldsymbol{\mathcal{F}}^-
=  \left(\mathcal{E}_0\thinspace, \thinspace\thinspace
\mathcal{\vec E} 
-\mathcal{\vec B}\thinspace\right).
\end {cases}
\label{eq:qt_em_fields_final}
\end{equation}

Next using (\ref{eq:qt_em_fields_final}), we derive two quaternion expressions for the Lorentz electromagnetic force,
\begin{equation}
\begin{cases}
\boldsymbol{{F}\thinspace^+}
=  q\left(\mathcal{E}_0\thinspace, \thinspace\thinspace
\mathcal{\vec E} 
+\mathcal{\vec B}\thinspace\right)

\\
\\
\boldsymbol{{F}\thinspace^-} 
=  q\left(\mathcal{E}_0\thinspace, \thinspace\thinspace
\mathcal{\vec E} 
-\mathcal{\vec B}\thinspace\right)
\end {cases}
\label{eq:qt_Lorentz_foce}
\end{equation}
where $q$ is a unit charge, and electromagnetic field components are given by (\ref{eq:qt_em_velocity_fields}).  While the first expression in (\ref{eq:qt_Lorentz_foce}) represents positive electric charges, we suggest that the second expression corresponds to negative electric charges due to the opposite effect of the magnetic field.

For the stationary case, $\vec v\sim 0$, the electromagnetic fields are,
\begin{equation}
\begin{cases}
\boldsymbol{\mathcal{F}^+} 
=\boldsymbol{\mathcal{F}^-}
=\boldsymbol{\mathcal E}
\simeq \left(-\dfrac {\partial \phi_0} {c \partial t_0} \thinspace, \thinspace
-\vec{\nabla}\phi_0  
\thinspace\right),
\\
\\
\boldsymbol{\mathcal{F}_c}
=\boldsymbol{\mathcal B}
\simeq  0 \thinspace,
\end {cases}
\label{eq:qt_unified_stationary_em}
\end{equation}
which further reduces to the classical field expression for the stationary electro-static field, where $\partial \phi_0 / \partial t_0 \sim 0$,

\begin{equation}
\begin{cases}
\boldsymbol{\mathcal{F}^+} 
=\boldsymbol{\mathcal{F}^-}
=\vec{\mathcal{E}}
\simeq \thinspace 
-\vec{\nabla}\phi_0 \thinspace , 

\\
\\
\vec{\mathcal{B}}
\simeq \vec 0 \thinspace.
\end {cases}
\label{eq:qt_classical_em}
\end{equation}

It seems that the quaternion form of electromagnetic interaction demonstrates a full quaternion electric field and a classical vector magnetic field. In addition, it predicts existence of positive and negative electric charges that propagate differently in the magnetic field, as expected from the Hall effect \cite{Lorentz2}.

Finally, we derive new form of Maxwell electromagnetic equations from (\ref{eq:qt_generic_matter_densities}) and  (\ref{eq:qt_em_fields}),

\begin{equation}
\begin{cases}
 {\rho_0}  
=  \vec{\nabla}\cdot\vec{\mathcal E}-\dfrac {\partial \mathcal{E}_0} {c\partial t_0} \thinspace,

\\
\\ 
 \dfrac {\vec j^+}  {c}
= 
-  \dfrac {\partial \vec{\mathcal E}} {c\partial t_0}
-\vec{\nabla} \mathcal{E}_0 
-\vec{\nabla}\times \vec{\mathcal B}+2\dfrac {\partial \vec{\mathcal B}} {c \thinspace \partial t_0}\thinspace,
\\
\\ 
 \dfrac {\vec j^-}  {c}
= 
-  \dfrac {\partial \vec{\mathcal E}} {c\partial t_0}
-\vec{\nabla} \mathcal{E}_0 
-\vec{\nabla}\times \vec{\mathcal B}-2\dfrac {\partial \vec{\mathcal B}} {c \thinspace \partial t_0}\thinspace,
\\
\\ 
 \dfrac {\vec j^\sim}  {c}
= 
-  \dfrac {\partial \vec{\mathcal E}} {c\partial t_0}
-\vec{\nabla} \mathcal{E}_0 
+\vec{\nabla}\times \vec{\mathcal B}\thinspace,
\\
\\ 

\vec{\nabla}\times \vec{\mathcal E}
+ \dfrac {\partial \vec{\mathcal B}} {c \thinspace \partial t_0}
=0\thinspace,
\\
\\
\vec{\nabla}\cdot\vec{\mathcal B} 
= 0\thinspace.
\end {cases}
\label{eq:qt_maxwel_new}
\end{equation}

It appears that the current density includes a gradient of a scalar function, $\vec{\nabla} \mathcal{E}_0$, which may correspond to the solid-state current given by the gradient of the quasi-Fermi potential. Also, four separate kinds of the charge density may correspond to the four different charge types in solid-state: electrons, holes, positive and negative ions.

\section{Quaternion Gravitational Fields and Matter Density} %

Next, let us assume that the gravitational field can be also described by a potential function in the quaternion form (\ref{eq:qt_potential}). Then, we  apply definitions of the  quaternion field components (\ref{eq:qt_fileds_define}) in order to derive two components of the gravitational field,
\begin{equation}
\begin{cases}
\boldsymbol{\mathcal{F}_a}
= \boldsymbol{\varGamma}
=   \left( \varGamma_0 \thinspace, \thinspace\thinspace
\mathcal{\vec \varGamma}\thinspace\right)\thinspace,
\\
\\
\boldsymbol{\mathcal{F}_c}
= \boldsymbol{\varOmega}
=  \left(0, \thinspace\thinspace
\mathcal{\vec \varOmega}\thinspace\right) \thinspace .
\end {cases}
\label{eq:qt_grav_define}
\end{equation}
where the components of the gravitational field include the novel scalar field, $\varGamma_0$, in addition to two vector fields, $\vec \varGamma$, and  $\vec \varOmega$. Now, we can calculate the field components from the potential using (\ref{eq:qt_unified_fields}),

\begin{equation}
\begin{cases}
\varGamma_0
= -\dfrac {\partial \phi_0} {c \partial t_0} 
+\thinspace\vec {\nabla} \cdot \vec{\phi}_v\thinspace,
\\
\\	
\vec \varGamma
=  -\vec{\nabla} \phi_0 -\dfrac {\partial \vec{\phi}_v} {c \partial t_0},
\\
\\
\vec \varOmega
= \thinspace\vec{\nabla}\times \vec{\phi}_v \thinspace .
\end{cases} 
\label{eq:qt_grav_fields}
\end{equation}
where, $\vec \varGamma$, and, $\vec \varOmega$, are gravitational equivalents of the electrical and magnetic fields. 

Using the velocity dependent vector potential, $\phi_v = (\vec v /c ) \phi$, we obtain,
\begin{equation}
\begin{cases}
\varGamma_0
= -\dfrac {\partial \phi_0} {c \partial t_0} +\phi \thinspace \vec {\nabla} \cdot \dfrac {\vec{v}} {c} + (\vec {\nabla}\phi)  \cdot \dfrac {\vec{v}} {c}\thinspace,
\\
\\	
\vec \varGamma
=\thinspace
 -\dfrac {\partial (\vec{v}\phi)} {c^2 \partial t_0}  -\vec{\nabla}\phi_0  
\thinspace,
\\
\\
\vec \varOmega
=  
\phi \left (\vec {\nabla} \times \dfrac {\vec{v}} {c}\right ) +(\vec {\nabla}\phi)  \times \dfrac {\vec{v}} {c}\thinspace.
\end{cases} 
\label{eq:qt_grav_velocity_fields}
\end{equation}

By applying (\ref{eq:qt_grav_define}) to the generic definition of the force fields (\ref{eq:qt_force total}), we obtain quaternion expressions for the total gravitational fields,

\begin{equation}
\begin{cases}
\boldsymbol{\mathcal{F}}^+
= \left(\varGamma_0\thinspace, \thinspace\thinspace
\vec \varGamma 
+\vec \varOmega\thinspace\right),

\\
\\
\boldsymbol{\mathcal{F}}^-
= \left(\varGamma_0\thinspace, \thinspace\thinspace
\vec \varGamma 
-\vec \varOmega\thinspace\right),
\end {cases}
\label{eq:qt_total_gravitational_field}
\end{equation}
which are of course equivalent to the general expressions (\ref{eq:qt_general_Lorentz_force}). The two expressions for the gravitational field differ by the direction of the torsion field, $\vec \varGamma$, similar to the effect of the magnetic field in the electromagnetic force. Therefore, we interpret the two field expressions as representation of two types of particle mass.

Then by using (\ref{eq:qt_total_gravitational_field}), we derive quaternion expressions for the gravitational forces for negative and positive masses respectively,

\begin{equation}
\begin{cases}
\boldsymbol{{F}\thinspace^+} 
=  m\left(\varGamma_0 \thinspace, \thinspace\thinspace
{\vec \varGamma} 
+{\vec \varOmega} \thinspace\right),
\\
\\
\boldsymbol{{F}\thinspace^-}
=  m\left(\varGamma_0 \thinspace, \thinspace\thinspace
{\vec \varGamma} 
-{\vec \varOmega} \thinspace\right),
\end {cases}
\label{eq:qt_Gravitational_force}
\end{equation}
where $m$ is a unit mass.

Assuming small variations of the gravitational potential with time, $\partial \phi_0 / \partial t_0 \sim 0$, and $\partial \vec \phi / \partial t_0 \sim 0$, we obtain an approximate form of the gravitational field,

\begin{equation}
\begin{cases}
\varGamma_0
\simeq  \phi \thinspace \vec {\nabla} \cdot \dfrac {\vec{v}} {c} + (\vec {\nabla}\phi)  \cdot \dfrac {\vec{v}} {c}\thinspace,
\\
\\	
\vec \varGamma
\simeq - \vec{\nabla} \phi_0 ,
\\
\\
\vec \varOmega
\simeq \phi \left (\vec {\nabla} \times \dfrac {\vec{v}} {c}\right ) +(\vec {\nabla}\phi)  \times \dfrac {\vec{v}} {c}\thinspace.
\end{cases} 
\label{eq:qt_grav_fields_approx}
\end{equation}
which is a new form of gravitational field expressions for slow varying fields.

For the stationary case, when, $\vec v\sim 0$, the field expressions reduce to,

\begin{equation}
\begin{cases}
\boldsymbol{\mathcal{F}^+} 
=\boldsymbol{\mathcal{F}^-}
=\vec{\varGamma}
\simeq \thinspace 
-\vec{\nabla}\phi_0 \thinspace ,

\\
\\
\vec{\varOmega}
\simeq \vec 0 \thinspace.
\end {cases}
\label{eq:qt_classical_gravitation}
\end{equation}
Thus we obtained quaternion expressions for the gravitational fields and forces similar to the electromagnetic expressions.

For the gravitational field, assume that, $-G\boldsymbol \rho$, replaces, $\boldsymbol \rho$, where $G$ is the gravitational constant. Let us assume that the gravitational matter density, $-  G\boldsymbol \rho$, can be expressed as the second derivative of the gravitational quaternion potential $\boldsymbol \phi$. Then, we derive from (\ref{eq:qt_generic_matter_densities}) and  (\ref{eq:qt_grav_fields}),

\begin{equation}
\begin{cases}
-  G\thinspace{\rho_0}  
=  \vec{\nabla}\cdot\vec{\varGamma}-\dfrac {\partial \varGamma_0} {c\partial t_0} 
\\
\\ 
-  G\thinspace\dfrac {\vec j^+}  {c}
= 
  -\dfrac {\partial \vec{\varGamma}} {c\partial t_0}
-\vec{\nabla} \varGamma_0 
-\vec{\nabla}\times \vec{\varOmega}+2\dfrac {\partial \vec{\varOmega}} {c \thinspace \partial t_0},
\\
\\ 
-  G\thinspace\dfrac {\vec j^-}  {c}
= 
- \dfrac {\partial \vec{\varGamma}} {c\partial t_0}
-\vec{\nabla} \varGamma_0 
-\vec{\nabla}\times \vec{\varOmega}-2\dfrac {\partial \vec{\varOmega}} {c \thinspace \partial t_0},
\\
\\ 
-  G\thinspace\dfrac {\vec j^\sim}  {c}
= 
- \dfrac {\partial \vec{\varGamma}} {c\partial t_0}
-\vec{\nabla} \varGamma_0 
+\vec{\nabla}\times \vec{\varOmega},
\\
\\ 
\vec{\nabla}\times \vec{\varGamma}
+ \dfrac {\partial \vec{\varOmega}} {c \thinspace \partial t_0}
=0,
\\
\\
\vec{\nabla}\cdot\vec{\varOmega} 
= 0.
\end {cases}
\label{eq:qt_new_gravitational_density}
\end{equation}

Thus, we derived the new matter density equations, which are very similar to the electromagnetic charge density expressions in the new form of Maxwell equations. Furthermore, we obtained additional density and current terms, resulting from the scalar fields, which may be responsible for the dark energy and matter.

\section{Conclusions} %
We introduced quaternion space-time and presented a framework for description of physical events using quaternion time intervals.  
We derived the quaternion form of the Lorentz time dilation and presented an intuitive physical interpretation of the time transformation between the source and observer reference frames. We showed that the resulting physical interpretation is inseparable from experimental measurements. Then, we used the quaternion algebra to derive quaternion calculus in the observer reference frame, by choosing the correct quaternion differentiation procedure. 
We applied the new derivative to the generalize potential function and suggested that the result can be interpreted as the quaternion force field, similar to the generalized Lorentz force. We repeated the differentiation procedure and assumed that the second derivative of the potential function can be interpreted as the generalized matter density. Then, we applied the first and second quaternion derivatives to electromagnetic and gravitational interactions. The novel terms in the field and matter density expressions result from the scalar fields and need further study and experimental verification.  Therefore, the new quaternion interpretations of force fields and matter density were derived directly from the definitions of quaternion derivatives and consequently present a unified mathematical framework, which may be used as the basis for the unified theory of space-time, matter, and reality.

\end{document}